\documentclass[journal,peerreview,nodraftcls]{IEEEtran}
\usepackage{amsmath,verbatim,amsthm,latexsym,amssymb}
\usepackage{graphicx,subfig}
\usepackage{enumerate}
\usepackage{textcomp}
\usepackage[hyphens]{url} 
\usepackage{hyperref}
\usepackage{setspace}
\usepackage{wrapfig}
\usepackage{booktabs} 

\newcommand{\mP}{\mathbb{P}}
\newcommand{\bfM}{\mathbf{M}}
\newcommand{\E}{\mathbb{E}}

\newcommand{\bSigma}{\boldsymbol{\Sigma}}

\newcommand{\bPsi}{\boldsymbol{\Psi}}

\newcommand{\bTheta}{\boldsymbol{\Theta}}

\newcommand{\bOmega}{\boldsymbol{\Omega}}

\newcommand{\bM}{\boldsymbol{M}}

\newcommand{\bI}{\boldsymbol{I}}

\newcommand{\bS}{\boldsymbol{S}}

\newcommand{\bX}{\boldsymbol{X}}

\newcommand{\bZ}{\boldsymbol{Z}}

\newcommand{\bOO}{\boldsymbol{O}}

\newcommand{\mN}{\mathcal N}

\newcommand{\argmax}{\operatornamewithlimits{argmax}}

\newcommand{\bit}{\begin{itemize}}
\newcommand{\eit}{\end{itemize}}
\newcommand{\ben}{\begin{enumerate}}
\newcommand{\een}{\end{enumerate}}
\newcommand{\beqn}{\begin{equation}}
\newcommand{\eeqn}{\end{equation}}
\newcommand{\bea}{\begin{eqnarray*}}
\newcommand{\eea}{\end{eqnarray*}}
\newcommand{\bpf}{\begin{proof}}
\newcommand{\epf}{\end{proof}\ms}
\newcommand{\ms}{\medskip}

\newtheorem{definition}{Definition}

  \usepackage{cite}

\begin{document}
\title{Classification with the matrix-variate-$t$ distribution}
\author{{Geoffrey Z. Thompson, Ranjan Maitra, William      Q. Meeker and Ashraf Bastawros}
  \IEEEcompsocitemizethanks{\IEEEcompsocthanksitem The authors are
    with Iowa State University. Email: \{gzt,maitra,wqmeeker,bastaw\}@iastate.edu.}
  \IEEEcompsocitemizethanks{This research was supported in part by the
      National Institute of Justice (NIJ) under Grants
      No. 2015-DN-BX-K056 and 2018-R2-CX-0034. The research of the
      second author was also supported in part by the National
      Institute of Biomedical Imaging and Bioengineering (NIBIB) 
of the National Institutes of Health (NIH) under Grant R21EB016212,
and the United States Department of Agriculture (USDA) National
Institute of Food and Agriculture (NIFA) Hatch project IOW03617.
The content of this paper is however solely the responsibility of the
authors and does not represent the official views of the NIJ, the
NIBIB, the NIH, the NIFA or the USDA.}}


\markboth{
}%
{Thompson \MakeLowercase{\textit{et al.}}:Classification with the Matrix-variate $t$-distribution}
\clearpage
\setcounter{page}{1}


\IEEEcompsoctitleabstractindextext{%

  \begin{abstract}
Matrix-variate distributions can intuitively model the
  dependence   structure of matrix-valued observations 
  that arise in applications with multivariate time
  series, spatio-temporal or repeated measures. 
  This paper develops an Expectation-Maximization algorithm for discriminant analysis and classification
with matrix-variate $t$-distributions. The methodology shows promise
on simulated datasets or when  applied to the forensic matching of
fractured surfaces or the classification of functional Magnetic Resonance,
satellite or hand gestures images. 
\end{abstract}

}
\maketitle

\IEEEdisplaynotcompsoctitleabstractindextext

  \section{Introduction}\label{introduction}
Matrix-variate distributions~\cite{gupta1999matrix} can 
conveniently model matrix-valued observations that arise, for instance, with
multivariate time series or spatial datasets or when we observe $p$-variate
responses at $q$ different settings, yielding a $p \times q$ matrix
of responses for  each unit of observation.  The matrix-variate normal
distribution 
(abbreviated in this paper by  MxVN) is also helpful for inference, but
is sometimes inadequate for modeling populations where
the matrix-variate-$t$ distribution (henceforth MxV$t$) may be a
better fit.     

There exist discriminant analysis and classification methods for 
MxVN~\cite{viroli2011, anderlucci2015} mixtures  
but for many 
applications, the MxV$t$
distribution may model each group better.
However, parameter estimation for the MxV$t$ distribution requires
special care because, unlike in the normal case, it can not be
viewed as simply a rearrangement of its
vector-multivariate cousin~\cite{dickey1967} 
for which several variants of
the Expectation-Maximization (EM) algorithm
exist~\cite{dempster1977maximum,meng1993,liurubin1994}.

This paper develops, in Section~\ref{methodology}, methodology for
parameter estimation in the MxV$t$ distribution and extends it to
discriminant analysis and classification using MxV$t$
mixtures. Our methods are evaluated on
simulated and real-life datasets in Section~\ref{performance}. This paper
concludes with some discussion. An online supplement explicitly
detailing the derivations of our algorithm, with sections
 referenced using the prefix ``S-'',
and an R~\cite{rproject} package
\href{http://www.github.com/gzt/MixMatrix}{\texttt{MixMatrix}}~\cite{thompson19} that 
implements the methodology are also included. 

\section{Methodology}
\label{methodology}
\subsection{Background and Preliminary Development}
\label{prelim}
\subsubsection{The Matrix-variate Normal Distribution}
\begin{definition}
  \label{defn:normal}
  A random matrix $\bX$ of $p$ rows and $q$ columns has the MxVN
distribution with parameters $\bfM, \bSigma$ and $\bOmega$ if it has
the probability density function (PDF) 
\begin{align*}f(\bX;\mathbf{M}, \bSigma, \bOmega) &= \frac{\exp\left( -\frac{1}{2} \, \mathrm{tr}\left[ \bOmega^{-1} (\bX - \mathbf{M})^{T} \bSigma^{-1} (\bX - \mathbf{M}) \right] \right)}{(2\pi)^{pq/2} |\bOmega|^{p/2} |\bSigma|^{q/2}}, \end{align*}
where $|\cdot |$ denotes the determinant, $\mathbf{M}$ is a
$p \times q$ matrix that is the mean of $\bX$,
and $\bSigma$ and $\bOmega>0$  describing
the covariances between, respectively, each of the $p$ rows and the
$q$ columns of $\bX$. We write $\bX\sim \mathcal N_{p,q}(\mathbf{M},
\bSigma, \bOmega)$. For identifiability, we set the first element of
$\bSigma$ to be unity.
\end{definition}
The MxVN  distribution can be considered, after rearranging into a
vector (denoted by vec($\bX$)), to be from a multivariate normal (MVN)
distribution with a Kronecker product covariance structure
\cite{gupta1999matrix}. So, if $\bX\sim \mathcal N_{p,q}(\mathbf{M},
\bSigma, \bOmega)$, then $\mathrm{vec}(\bX) \sim
\mN_{pq}(\mathrm{vec}(\mathbf{M}), \bOmega \otimes \bSigma).$ This
reformulation allows us to readily obtain the maximum likelihood (ML)
estimates. If $\bX_i$, $i = 1, 2, \ldots, n$ are independent
identically distributed (IID) random matrices from the $\mathcal N_{p,q}(\mathbf{M},
\bSigma, \bOmega)$,  then $\mathbf{M}, \bSigma,$ and $\bOmega$ have
ML estimators 
$\widehat{\mathbf{M}}  =  {n}^{-1} \sum_{i = 1}^n \bX_i, 
    \widehat{\bSigma}  =  (np)^{-1} \sum_{i = 1}^n (\bX -
    \widehat{\mathbf{M}}) \widehat{\bOmega} ^{-1}(\bX -
    \widehat{\mathbf{M}})^T$, and $\widehat{\bOmega}   = (nq)^{-1}
    \sum_{i = 1}^n (\bX - \widehat{\mathbf{M}})^T \widehat{\bSigma}
    ^{-1}(\bX - \widehat{\mathbf{M}})$ under the constraint for identifiability that the
    $(1,1)$th element of $\widehat{\bSigma}$ is set to unity.
    The matrices $\widehat{\bSigma}$ and $\widehat{\bOmega}$ are obtained
    iteratively after initialization with any positive definite matrices,
    by default using the identity matrix $\bI$. 
    The ML estimates exist and are unique almost surely if $n > p/q + q/p + 2$
    \cite{SOLOVEYCHIK201692}. 
    \subsubsection{The Matrix-variate $t$-distribution}
    \label{matrix-variate-t-distribution}
    \begin{definition}
      \label{defn:t}
      A random $p \times q$ matrix $\bX$ has a MxV$t$ distribution
      with parameters
      $(\bfM,\bSigma,\bOmega$) of similar order as in
      Definition~\ref{defn:normal} (with $\bSigma$ and $\bOmega > 0$) and degrees of freedom (df) $\nu\geq1$  if its PDF is  
\begin{equation*}f(\bX;\nu, \mathbf{M}, \bSigma, \bOmega) =
             {\frac{\Gamma _{p}\left({\frac  {\nu +p+q-1}{2}}\right)}{(\pi )^{{\frac  {pq}{2}}}\Gamma _{p}\left({\frac  {\nu +p-1}{2}}\right)}}|
               \bOmega |^{{-{\frac  {p}{2}}}}|\bSigma |^{{-{\frac  {q}{2}}}} 
              \left|{\mathbf  {I}}_{p}+  \bSigma^{{-1}}(\bX-{\mathbf{M}})\bOmega^{{-1}}(\bX-{\mathbf  {M}})^{{{\rm {T}}}}\right|^{{-{\frac  {\nu +p+q-1}{2}}}}.\end{equation*}
We use the notation   $\bX\sim
t_{p,q}(\nu,\bfM,\bSigma,\bOmega)$  to indicate that $\bX$ has
this density.
\end{definition}
          \paragraph{Properties:} We mention some properties of the MxVt
          distribution relevant to this paper.
          \begin{enumerate}
            \item \label{prop1}
              For $p = 1$ and $\bSigma \equiv \nu$ (or $q = 1$ and
              $\bOmega \equiv \nu$), the MxV$t$ distribution reduces to its
              vector-multivariate $t$ (MVT) cousin. However, this
              reduction does not generally hold so additional
              development is needed for inference. We provide methods
              to do so in the next section.
\item \label{prop2} Let the random matrix $\bS\sim  {\mathcal W}_p(\nu + p -1, \bSigma^{-1})$,
  where ${\mathcal W}_p(\kappa,\bPsi)$ is the $p\times p$-dimensional
  Wishart distribution with d.f. $\kappa$ and scale matrix $\bPsi$. If
  $\bX\mid \bS\sim {\mathcal N}_{p,q}(\bfM,\bS^{-1}, \bOmega)$,
  then  $\bX\sim t_{p,q}(\nu, \bfM, \bSigma, \bOmega)$~\cite[see][p. 135]{gupta1999matrix}. 
Further, $\bS\mid\bX \sim {\mathcal W}_p(\nu+p+q-1, [(\bX-\mathbf{M})\bOmega^{-1}(\bX-\mathbf{M})^T + \bSigma]^{-1})$~\cite{iranmanesh2010conditional}.
\end{enumerate}

\subsection{ML Estimation  of the MxV$t$ parameters}
\label{ecme.t}
 The MxV$t$ distribution does not
have closed-form ML estimators so we provide an Expectation/Conditional
Maximization Either (ECME) algorithm~\cite{liurubin1994} in a manner that
is similar to that used to find ML parameter estimates in the MVT distribution,
with the main contribution being the extension to the matrix variate case by deriving
the estimates in terms of a matrix variate normal mixture with a Wishart distribution
rather than a multivariate normal mixture with a chi-squared distribution.

Let $\bX_i ,i=1,2,\ldots,n$ be IID
realizations from $t_{p,q}(\nu,\bfM,\bSigma,\bOmega)$.
Write $\bTheta\equiv\{\nu,\bfM,\bSigma,\bOmega\}$. For each
$i=1,2,\ldots,n$, let $\bS_i$ be (unobserved) Wishart-distributed
random matrices that are as per Property~\ref{prop2}.

From the detailed development and derivations provided in the
supplement Section~\ref{mle-of-parameters-m-omega-sigma-and-nu-using-ecme},
we get the expectation step (E-step) updates 
at the 
current value  $\bTheta^{(t)}$ of $\bTheta$ by taking the expected values
of the $\bS_i$ given the current value of  $\bTheta^{(t)}$:
\begin{equation*}
  \begin{split}
\bS_{i}^{(t+1)} \doteq\E_{\Theta^{(t)}}(\bS_i | \bX_i) &= (\nu^{(t)}+p+q-1)[(\bX_i-\mathbf{M}^{(t)})\bOmega^{(t)^{-1 }}(\bX_i-\mathbf{M}^{(t)})^T + \bSigma^{(t)}]^{-1} \\
\E_{\Theta^{(t)}}\left(\log| \bS_i| \big| \bX_i \right)   &=\psi_{p}\left({\frac {\nu^{(t)} + p + q -1}{2}}\right) + p \log 2 + \log \left|\frac{\bS_{i}^{(t+1)}}{\nu^{(t)} + p + q -1}\right|
\end{split}
\end{equation*}
where $\psi_p(\cdot)$ is the $p$-variate digamma function -- 
that is, $\psi_p(x) = \frac d{dx} \log\Gamma_p(x)$. Further
computational and notational reductions are possible by defining and storing the
updates in terms of the expected sufficient statistics
\begin{equation*}
  \begin{split}
    \bS_{S}^{(t+1)} &\doteq \sum_{i=1}^n \bS_{i}^{(t+1)},\\
    \bS_{SX}^{(t+1)} & \doteq  \sum_{i=1}^n \E_{\Theta^{(t)}} (\bS_i\bX_i|\bX_i)
=\sum_{i=1}^n \bS_{i}^{(t+1)}\bX_i, \\
\bS_{XSX}^{(t+1)} &\doteq \sum_{i=1}^n\E_{\Theta^{(t)}}
(\bX_i^T\bS_i\bX_i|\bX_i) = \sum_{i=1}^n
\bX_i^T\bS_{i}^{(t+1)}\bX_i,\\
\bS_{|S|}^{(t+1)} & \doteq \E_{\Theta^{(t)}}  \left[\sum_{i = 1}^n \log| \bS_i|  \bigg| \bX_i \right]. 
\end{split}
\end{equation*}
These statistics can be expressed with $(\nu^{(t)} + p + q -1)$
factored out, and for convenience may be computed and stored as such
when $\nu$  needs to be estimated. These quantities can be computed
in $\bOO(npq^2) + \bOO(np^2q) + \bOO(np^3)$ flops.

The M-step updates are split into two conditional maximization steps,
one updating  $(\bfM,\bSigma,\bOmega)$ and one updating $\nu$.
The first step is conceptually immediate and maximizes Equation
\eqref{qfn} in the supplement with respect to $\bTheta$ to yield
$\bTheta^{(t+1)}$, with updates of $(\bfM,\bSigma,\bOmega)$ given
 $\nu^{(t)}$ as follows:
\begin{align*}
\widehat{\mathbf{M}}^{(t+1)} &= \left(\sum_{i = 1}^n\bS_i^{(t+1)}\right)^{-1}\sum_{i=1}^n\bS_i\bX_i  = {\bS_{S}^{(t+1)}}^{-1} \bS_{SX}^{(t+1)} \\
  \widehat{\bOmega}^{(t+1)} &= \frac{1}{np} \sum_{i = 1}^n (\bX_i - \widehat{\mathbf{M}}^{(t)})^T \bS_i^{(t+1)} (\bX_i - \widehat{\mathbf{M}}^{(t)}) =\frac{1}{np} \left(\bS_{XSX}^{(t+1)} - {\bS_{SX}^{(t+1)}}^T {\bS_{S}^{(t+1)}}^{-1}{\bS_{SX}^{(t+1)}}  \right) \\
 {\widehat{\bSigma}_{(t+1)}}^{-1} &= \frac{1}{n(\nu^{(t)}+p-1)}\sum_{i = 1}^n \bS_i^{(t+1)} = \frac{\bS_S^{(t+1)}}{n(\nu^{(t)}+p-1)}.
\end{align*}
 These quantities can be computed
 in $\bOO(pq^2) + \bOO(p^2q) + \bOO(p^3)$ flops. As discussed in
 Section~\ref{s-em}, if the previous $\bSigma^{(t)}$ and $\bOmega^{(t)}$
 were positive definite, then  the updates exist and $\widehat{\bSigma}$ is positive definite,
 though the necessary or sufficient conditions for $\widehat{\bOmega}$ to be positive definite (a.s.)
 are not known.
  
 The conditional maximization of $\nu$ given
$(\bfM^{(t+1)},\bSigma^{(t+1)},\bOmega^{(t+1)})$ can be sped up substantially
by maximizing it instead over the observed log-likelihood function given 
$(\bfM^{(t+1)},\bSigma^{(t+1)},\bOmega^{(t+1)})$. We get the ML estimating
equation (MLEE):
\begin{equation}
 n \frac{d}{d\nu}\log \Gamma_p ((\nu+p-1)/2) - \frac{1}{2} (\bS_{|S|}
 - np\log2 + n \log |\widehat{\bSigma}|) = 0.
 \label{mleqn}
\end{equation}
Writing $\kappa = \nu +p + q -1$ for notational compactness, we have,
\begin{align}
  \begin{split}
 0 &= n \psi_p((\nu+p-1)/2) -  \left\{n\psi_{p}\left({\frac {\kappa}{2}}\right)  +\sum_{i=1}^n \log \left|\frac{\bS_{i}^{(t+1)}}{\kappa}\right| - n \log \left|\frac{\bS_S^{(t+1)}}{n(\nu+p-1)}\right|\right\} \\*
 &=  \psi_p((\nu+p-1)/2) - \left\{\psi_{p}\left({\frac
   {\kappa}{2}}\right)  +\frac{1}{n}\sum_{i=1}^n \log
 \left|\bZ_{i}^{(t+1)}\right| + p \log \frac{n(\nu + p -1)}{\kappa} -
 \log \left|\bZ_S^{(t+1)}\right|\right\},\end{split}
 \label{eqn:ecme}\end{align}
where $\bZ_*^{(t+1)}$ is the appropriate $\bS_*^{(t+1)}$ statistic with
$(\nu^{(t)} + p + q -1)$ factored out. 
The MLEE can be solved using a one-dimensional search, yielding an
ECME algorithm with the  steps: 
\begin{enumerate}
\def\labelenumi{\arabic{enumi}.}
\item
  {\bf E-step:} Update $\bS_i$ weights and statistics based on
  $\Theta^{(t)}$ and $\bX_i$.
\item
  {\bf CME-step:} Update
  $\Theta_1^{(t+1)} = (\mathbf{M}^{(t+1)}, \bSigma^{(t+1)}, \bOmega^{(t+1)})$.
\item
  {\bf CME-step:} Update $\Theta_2^{(t+1)} = \nu^{(t+1)}$ using the observed
  log-likelihood given $(\mathbf{M}^{(t+1)},\bSigma^{(t+1)},\bOmega^{(t+1)})$
  .
\end{enumerate}
Repeat these steps until convergence.

As explained in Section~\ref{s-em}. each iteration of this algorithm takes
$\bOO(npq^2) + \bOO(np^2q) + \bOO(np^3)$ flops in addition to the
number of iterations required to estimate $\nu$ in 
the second CME step. This suggests that the orientation of the
matrices should be chosen such that the row dimension $p < q$, a
suggestion that is also experimentally verified in Section~\ref{simulation-study}.

We conclude here by noting that restrictions on the parameter
set~\cite{ROY2005462}, such as imposing an $m$th-order auto-regressive structure
(AR($m$)) on either  $\bSigma$ or $\bOmega$ or both as in our
applications in Section~\ref{performance} can be easily incorporated
within our algorithm (see
Section~\ref{fitting-with-restrictions-on-the-parameters}). 
  \subsection{Discrimination and Classification}\label{classification}
  Linear (LDA) and Quadratic Discriminant Analysis (QDA) for
  matrix-variate populations follow a similar approach as for the
  multivariate case, with the MxVN (but not MxV$t$) cases affording substantial
  reductions in the computations. We provide here the general
  framework for matrix-variate distributions and then discuss
  reductions for the cases of the MxVN models.

Suppose that there are two populations   $\pi_1$ and $\pi_2$, with
prior probabilities $\eta_1$ and $\eta_2$ for an observation belonging
to either. 
Let $\mP(1|2)$ be the probability of
classifying a member of  $\pi_2$ to $\pi_1$
(and vice versa). As usual, the \emph{total probability of
  misclassification} (TPM) is defined to be $\mP(2|1)\eta_1
+ \mP(1|2)\eta_2$. A Bayes optimal classification rule that minimizes
the TPM assigns a matrix-valued observation $\bX$ to $\pi_1$ if $
\frac{f_1(\bX)}{f_2(\bX)} \geq \frac{ \eta_2}{\eta_1}$, where
$f_i(\bX)$ is the PDF for group $\pi_i$ 
evaluated at $\bX$~\cite{AndersonBahadur}. The  classification rule can be easily extended to
the case when there are $G$ groups $\pi_1,\pi_2,\ldots,\pi_G$, each
with prior probabilities of membership $\eta_1, \eta_2, \ldots,
\eta_G$ and densities $f_1, f_2, \ldots, f_g$. Then the Bayes optimal
classification for a matrix-valued observation $\bX$ is $\argmax_{i
  \in \{1, 2, \ldots, g \}}R_i( \bX)$, where the cost function $R_i( \bX)$
is defined as $ \log \eta_i f_i(\bX)$. 

Unlike for the MxV$t$ distributions, the MxVN case has closed-form solutions
analogous to that of LDA or QDA in multivariate statistics.
For the MxVN populations, the closed-form classification rule
 assigns $\bX$ to the $g$th group where
$g=\argmax_{i=1,2,\ldots,G}R_i(\bX)$, with 
\begin{equation}
  \begin{split}
R_i(\bX) &= \mathrm{trace}\left\{ -\frac{1}{2}(\bOmega_i^{-1} \bX^{T} \bSigma_i^{-1} \bX )  +\bOmega_i^{-1} \mathbf{M}_i^{T} \bSigma_i^{-1}\bX 
  -\frac{1}{2}\bOmega_i^{-1} \mathbf{M}_i^{T} \bSigma_i^{-1}
           \mathbf{M}_i  \right\} \\
  & \qquad-\frac{1}{2}(p \log|\bSigma_i|+ q\log|\bOmega_i| )   .
\end{split}
\label{g.qda}
\end{equation}
The first and last term disappear when the $G$ MxVN populations have common
covariances, yielding a linear decision rule. 
Many adaptations~\cite{bf88260936d14048a3d6bc78fc588842,inoue2011, Li2005SLD, Lu2009UMD,mahanta2014ranking,    yan2007multilinear, NIPS2004_2547,zhao2012separable, zheng20081d, doi:10.1080/10618600.2018.1476249}  of LDA exist for homogeneous MxVN
 populations, but our development provides a natural and direct approach that is
 also flexible enough to include a range of assumptions. Assuming homogeneity
 does not yield a linear rule for MxV$t$
 populations where we still get a quadratic rule. Finally, in all
 cases, the parameters in $R(\bX)$ can be estimated using ML on the
 training set (with the ECME methodology of Section~\ref{ecme.t} for MxV$t$
 populations) and incorporated into the decision rule.
\section{Performance Evaluations}\label{performance}

This section evaluates performance of the ECME algorithm in
recovering the MxV$t$ parameters and also classification performance
of our methodology on some real-life datasets.

\subsection{Simulation Study}\label{simulation-study}

\begin{wrapfigure}{r}{0.5\textwidth}
  \vspace{-.4in}
     \centering
    \mbox{
      \subfloat[]{\includegraphics[width=.5\textwidth]{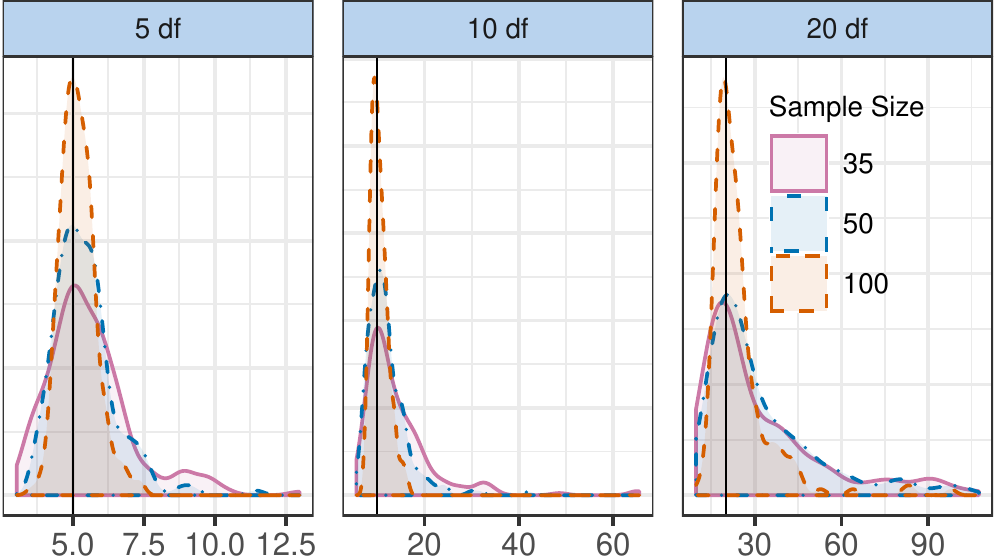}}}
    \mbox{
  \subfloat[]{\small
    \begin{tabular}{llcrrrr}
      \toprule
  $\nu$& $n$ & Range & Median & Mean & SD \\  \midrule
      5  & 35 &  $(2.62 , 15.12)$ & 5.24 & 5.45 & 1.43 \\ 
         & 50 & $(3.46 , 12.78)$ & 5.32 & 5.45 & 1.28 \\ 
         & 100 &  $(3.76,  7.01)$ & 5.14 & 5.18 & 0.56 \\ 
      10 & 35 & $(5.41 , 395.12)$ & 11.57 & 16.28 & 28.26 \\ 
         & 50 & $(5.28 , 106.72)$ & 10.44 & 11.94 & 7.91 \\ 
         & 100 & $(6.99 , 18.10)$ & 10.19 & 10.56 & 1.84 \\ 
      20 & 35 & $(9.93 , 999.83)$ & 29.94 & 89.01 & 149.12 \\
         & 50 & $(11.38, 495.67)$ & 24.45 & 46.97 & 73.09 \\ 
         & 100 &  $(12.68 , 147.67)$ & 21.98 & 25.52 & 14.91 \\ 
  \bottomrule
\end{tabular}
}}
\caption{(a) Density plots and (b) numerical summaries of  $\hat\nu$
  for datasets of size $n=35, 50, 100$ with true $\nu=5, 10, 20$
  (vertical line).\vspace{-0.2in} }
\label{fig:df}
\vspace{-.0in}
\end{wrapfigure}

Our simulation study generated 200 datasets from the
$t_{5,3}(\nu,\bfM,\bSigma,\bOmega)$ distribution with $\nu=5,10,20$ and $n\in
\{35, 50, 100\}$, with the smallest $n$ chosen to be larger than the 
number of parameters to be estimated, which was also large enough
for all but one of the simulations to converge,  and the larger sample sizes
were chosen to give an idea of consistency of parameter estimation. 
The ECME algorithm in Section~\ref{ecme.t}, with unconstrained
$(\bfM,\bSigma,\bOmega)$ was used to estimate the parameters.
Figure~\ref{fig:df} summarizes the estimated $\hat\nu$ over the 200
samples for each $\nu$. (We constrain $\hat\nu$ to be in
$(2,1000)$ in the Either step of the ECME algorithm.)

As expected, higher $n$ improves both accuracy and precision of the
estimates. For all nine cases, the peak of the distribution of
$\hat\nu$ was close to the true $\nu$ value. Lower values for $\nu$
were more easily estimated in the sense that for any $n$, the
$\hat{\nu}$ values are closer to the true 
$\nu$ (Figure~\ref{fig:df}). This may be because for larger true $\nu$, the
distributions are similar in a wider range (after all, as
$\nu\rightarrow\infty$, the distribution reduces to the MxVN).
For one aberrant sample with $\nu = 20$ and $n=35$, the optimizer attained the
upper bound and did not converge to an interior point. On the whole,
however, our simulation results indicate good performance of the ECME
algorithm in recovering the MxV$t$ parameters. Additional information
about convergence and the recovery of the center and scatter parameters is contained
in Section~\ref{supp:simulation-study}.

\begin{wrapfigure}{r}{.5\textwidth}
  \centering
  \vspace{-.4in}
\includegraphics[width = .5\textwidth]{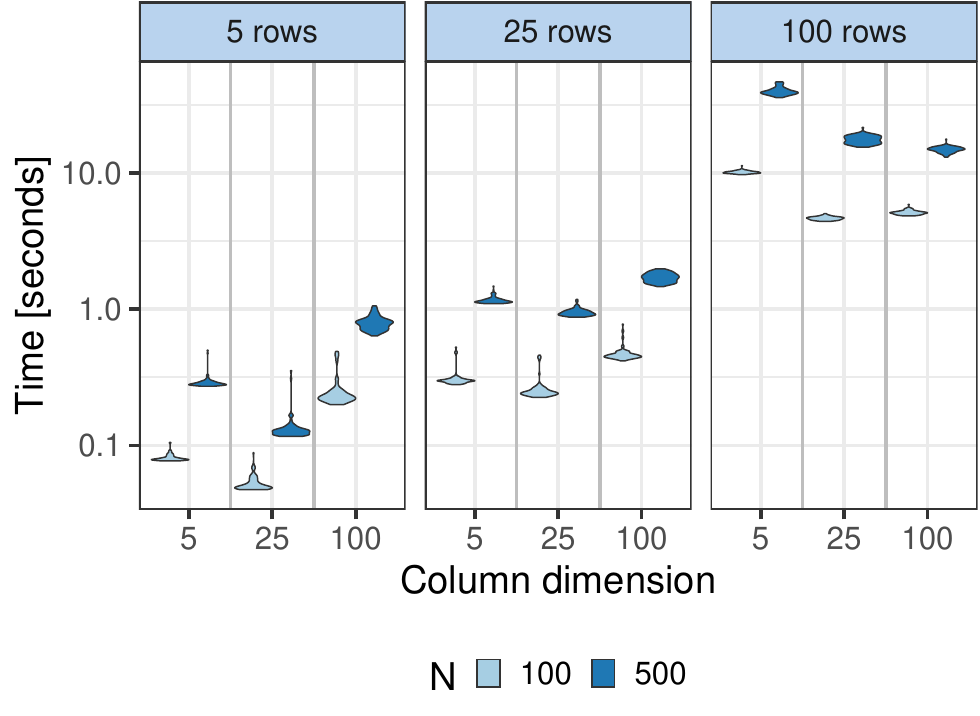}
\vspace{-0.2in}
\singlespace \caption{Run times for 100 repetitions
  of the proposed MxV$t$ estimation procedure for $p$= 5, 25, 100,
  $q$ = 5, 25, 100, and $N$ = 100, 500 with $\nu = 5$.
  }\label{fig:timing}
\vspace{-0.7in}
\end{wrapfigure}

In another simulation study, the speed of the algorithm was demonstrated
for $p = 5,\, 25,\, 100$,  $q = 5,\, 25,\, 100$, and $N = 100,\, 500$ with
$0$ mean and identity spread parameters for $\nu = 5$. Figure~\ref{fig:timing}
summarizes the results of the simulation. As the derivation in
Section~\ref{ecme.t} suggests, the row dimension $p$ dominates 
when determining the speed of the computation. Increasing column dimension $q$
for a given row dimension $p$ seems to hasten convergence to some extent,
likely because they act to effectively increase the sample size, which
reduces the number of iterations the algorithm needs to run.
This suggests the orientation should be chosen so that $p < q$.

Section~\ref{supp:simulation-study} also details results from a
a simulation study showing the performance of the method
when the type of model (MxV$t$ or MxVN) or degrees of freedom $\nu$ are
misspecified. We see that models that are fit with $\nu$ closer to the
true value perform better than those that do not. Also, as expected,
the performance using MxV$t$ more closely approaches that under the
MxVN model as $\nu$ increases. 

\subsection{Classification Examples}
We evaluate  MxV$t$  classification 
and discrimination on four different datasets.
\subsubsection{Matching Fractured Surfaces}\label{matching-of-fracture-surfaces}
Our first example is on the potential ability of our classification
algorithm to distinguish between pairs of fractured surfaces into
matches or non-matches, with implications in forensics to decide on, say, whether a knife blade fragment found at a
crime scene 
is a match to something that visually appears to be the remainder
of the blade. Because of the novelty of this application, we discuss
it at some length here. Our 
investigation is a formal proof-of-concept conducted in the lab
where a set of 
38 stainless steel knives had their blades broken under similar
conditions, resulting in each of them having a base and a tip. The
cross-sectional fractured surfaces were then scanned using a standard non-contact 3D optical interferometer 
 at 9 regularly-spaced locations to get 9 successive
$1024\times 1024$ images (with 75\% overlap, in order to get a
reasonable number of replications while also imaging the entire length
of the exposed surfaces.

\begin{wrapfigure}{r}{.5\textwidth}
  \centering
\includegraphics[width = .5\textwidth]{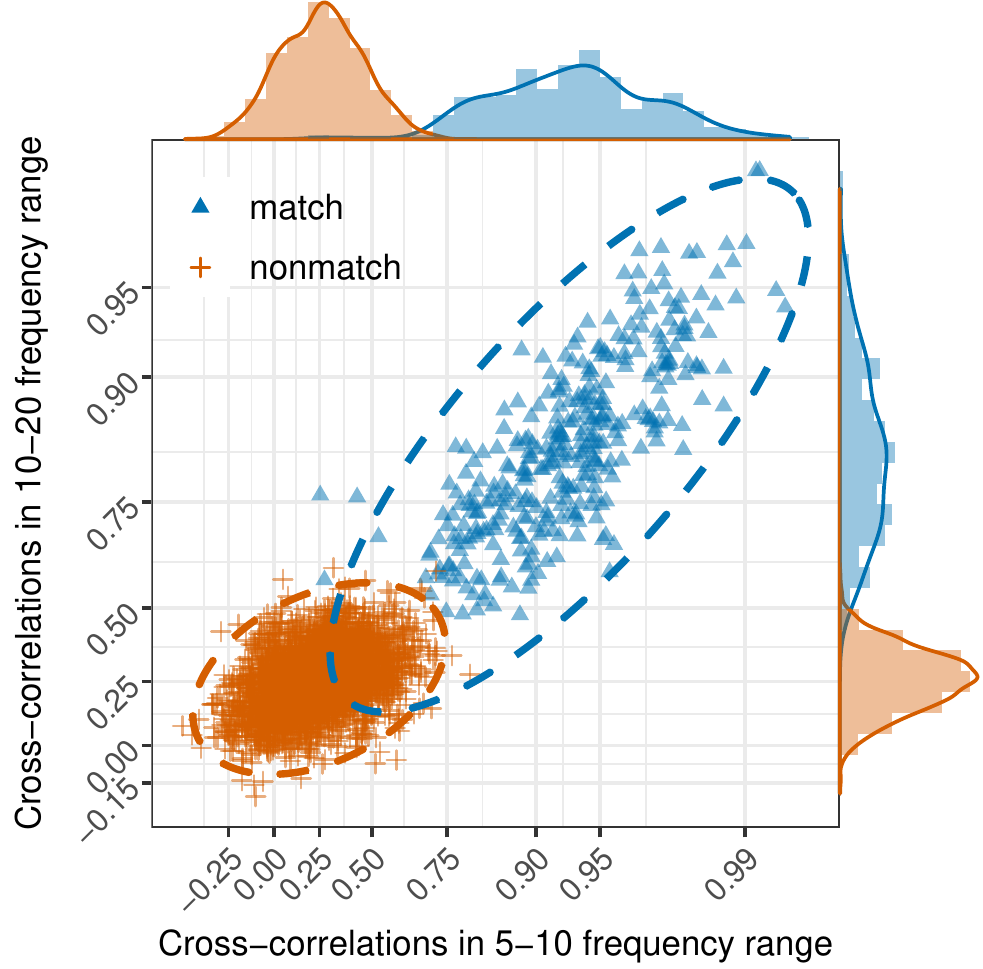}
\vspace{0.0in}
\singlespace \caption{Cross-correlations for individual images,
    along with 99\% confidence ellipses under bivariate normal
    assumptions, in known matching (KM) and known non-matching (KNM) surfaces on
    Fisher-transformed axes. 
    Known matches and known non-matches     can be distinguished, but
    not perfectly, in this example by     features in these two
    frequency   ranges. 
  }\label{fig:knifecorr}
\end{wrapfigure}
Cross-correlations between matching knife
base-tip image pairs in the 5-10 $\mu m^{-1}$ and 10-20
$\mu m^{-1}$ two-dimensional (2D) Fourier frequencies were computed,
yielding, for each knife, a
$2\times 9$ matrix of measurements describing the similarity of the
base of the knife to the tip (2 measured cross-correlations per image
and 9 images). Similar cross-correlations between all
possible knife base-tip pairs (regardless of origin) yielded a
sample from the population of similarity matrices coming from known
matching (KM) and known non-matching (KNM) base-tip pairs.

Figure~\ref{fig:knifecorr} shows the scatterplot of the Fisher's
$Z$-transformed~\cite{Fisher} cross-correlation data to be fairly
elliptical.  The two classes are almost but not completely
separated when looking at individual image pairs. Classification using
only one pair of images per surface rather than a set of 
multiple images is potentially ambiguous. We remove this
potential ambiguity by considering multiple images on each 
surface. These multiple sets of images on each knife are not
independent and have a natural multivariate repeated measures ({\em
  i.e.} matrix-variate) structure because of the 75\% overlap between
successive images so a model incorporating such structure 
may improve  classification accuracy.

We model each match/non-match
dataset in terms of the MxV$t$ distribution with group-specific 
mean matrix and matrix dispersion structures, with 
an AR(1) correlation structure for the Fourier domain correlations  at
the same frequency band between successive (overlapping) image
pairs. The AR(1) structure is appropriate because of the overlap
between successive images: this correlation structure also has
the best Bayesian Information Criterion (BIC) among the correlation
structures tested on the data~\cite{schwarz1978}. The mean across the
images for each frequency band was constrained to be constant.  
Because there are only 9-10 observations for the cases where
the knife tip-base have the same origin, 
we forgo estimating $\nu$ and instead investigate classification with
the MxV$t$ distribution with $\nu=5$ and $\nu=10$ (in addition to the MxVN). 

\begin{wrapfigure}{r}{.5\textwidth}
  \centering
  \vspace{-0.0in}
  \includegraphics[width = .5\textwidth]{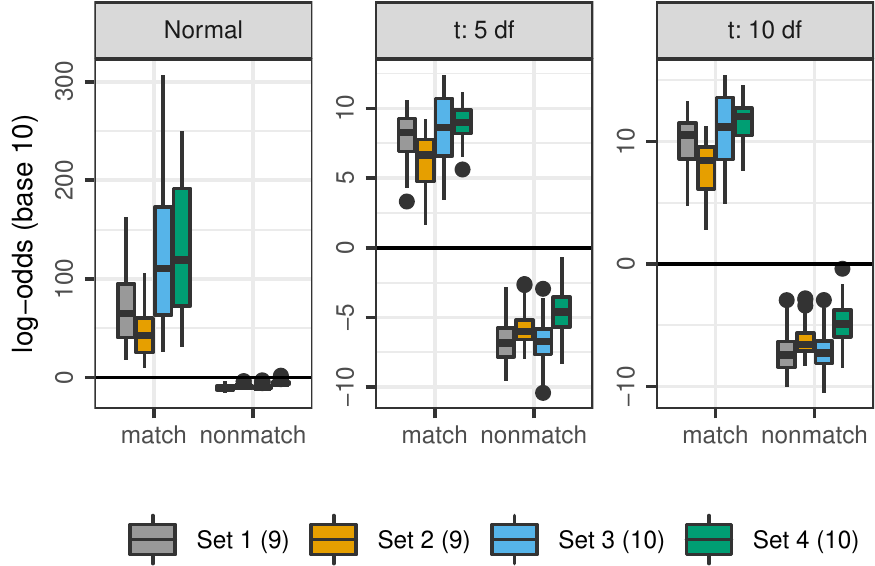}
\singlespace \caption{Positive results indicate match is
more probable than non-match. There are 9 known matches and 72 
known non-matches in Set 1 and Set 2 and 10 known matches and 90 
known non-matches in Sets 3 and 4. The results are from training on the indicated set
and testing on all four sets.
\vspace{-0.0in}}\label{fig:knifeclass}
\end{wrapfigure}

Figure~\ref{fig:knifeclass} displays the distribution of the log-odds
of being a match for the models based on each of the four training
sets. The models trained on each 
set were then tested on the data from all four sets of surfaces. In this figure,
a positive log-odds indicates a higher probability of being a KM and a negative
log-odds indicates a higher probability of being a KNM.
With equal priors, there is a 0\% false exclusion (false negative) rate and a 0.003\% false
identification (false positive) rate (1 FP). The only FP is from the MxVN model, which
is also overly confident about the matches it produces. It predicts
some surfaces being a match with log-odds greater than 200, which is
extremely implausible. The MxV$t$ distribution accounts for uncertainty
better and results in more plausible log-odds ratios. This is because
the normal distribution is much more thin in its tails than the
$t$-distribution, and increasing the dimensionality as occurs in the matrix variate
case multiplies this effect. This means the MxVN will penalize observations
far from the center of the class more than the MxV$t$. Perfect discrimination
is attained with MxV$t$ for all four training sets, suggesting that the results
generalize well to out-of-sample data despite 
the relatively small sample size.
For comparison, we also obtained predictions using the penalized likelihood
approach of \cite{doi:10.1080/10618600.2018.1476249} which works only
when at least two sets of knives are used as training sets, 
two sets as a validation set for the tuning parameters,
and the rest as the test set. We were able to obtain perfect classification
for all permutations of the six sets when an appropriate grid of tuning
parameters is used (two additional sets of images taken from one set of knives
were used to make a total of six sets). However, the method forced the scatter matrices
to be diagonal, which is unlikely to be reasonable given the 75\%
overlap between successive images.
\subsubsection{Finger-tapping Experiment}\label{fmri}
\cite{doi:10.1002/mrm.10191} provided 12 functional Magnetic
Resonance Imaging (fMRI) scans of the brain of a right-hand dominant
male subject during a right-hand finger-thumb opposition activity and
12 similar scans using the left-hand, with each pair of scan collected at
regular intervals over a 2-month period.  
We restrict attention to the 20th slice of the image volume, with
$128\times 128$ pixels, that 
previous work~\cite{maitra09b,maitra10} indicated as adequate to
distinguish activation between the left- and right-hand finger-tapping. 
With only 12 observations per class, we are limited in the types of
correlation matrices that we may consider, 
so we selected a $20 \times 20$ section of the 20th slice having the
left-topmost pixel at $(33,67)$, which was the $20 \times 20$ section
of the slice displaying the highest average activation in the
left-hand activation images as determined by~\cite{fastfmri}'s
FAST-fMRI algorithm. We then trained and tested the classifiers using the
leave-one-out method 
with  an AR(1) covariance structure and a compound symmetry covariance
structure in the MxVN and MxV$t$  distributions with $\nu=5$ or
$\nu=10$ (for the MxV$t$). 
The BIC on the fitted models indicated that a compound symmetry covariance
structure was the best model. In all cases, except that of the MxV$t$
distribution with $\nu = 10$ and an 
AR(1) covariance structure, 23 out 24 images were correctly
classified. The one mislabeled case was the same one that was
previously identified by~\cite{maitra10} as 
an outlier. Using the MxV$t$ distribution with $\nu=10$ had
one more misclassification. The number of cases for this reduced
dataset is not enough for {\tt MatrixLDA} to estimate the correlation
structure so we forgo that comparison here.
\subsubsection{Landsat Satellite Data}\label{landsat}
Multi-spectral satellite imagery allows for multiple observations over
a spatial grid, yielding matrix-valued observations. 
We examine a set of satellite images 
\cite{Dua:2019} that are 
in two visible and two infrared bands. The subset of images under consideration~\cite{Viroliclass2011} consists of a
training and a test set of $3 \times 3$ pixel 
segments labeled according to the terrain type (961 gray soil, 415 damp gray
soil, and 470 soil with vegetation stubble segments, for 1846 total observations
in the training set and 397, 311, 237, and 845 total in the test set) of their middle pixel.
Each observation, then,
is a 9-pixel segment with a label according to soil type, and the problem
is to predict the soil type from the pixel values.    
Regarding  the data as a $4 \times 9$ matrix and with an MxVN classifier
and unconstrained covariance matrices 
yielded an error
rate of 0.116~\cite{Viroliclass2011}, 
while {\tt MatrixLDA} with tuning parameters selected by 5-fold CV
\cite{doi:10.1080/10618600.2018.1476249} yielded a 0.118 error
rate. Our MxVN and  MxV$t$ models (the latter with $\nu=10$ and 20)
with unconstrained 
covariance matrices and prior probabilities equal to the class
representation in the training set yielded error rates
of 0.126, 0.116, and 0.109, in line with previous results. BIC indicated
that using unconstrained covariance matrices and means constrained to be equal
within rows as a better model, with error rates of 0.123, 0.121, and 0.107.

  \subsubsection{Cambridge Hand Gestures Data}\label{gestures}
We tested our method using leave-one-out cross-validation (LOOCV) on the set
of 80 images extracted from the Cambridge hand gestures database \cite{Kim2007TensorCC}
as processed by \cite{doi:10.1080/10618600.2018.1476249} into $80 \times 60$
pixel gray-scale images. There are
four classes in this problem: the images show a hand gesture in
one of two shapes and one of two orientations: in each image, the hand is either in a flat or ``V'' shape and is located either
in the center of the image or to the left side of the image. We fit models
with an AR(1) structure on both dimensions, compound symmetric structure on both,
and an unconstrained covariance structure, with 5 and 10 degrees of freedom
for the MxV$t$ distribution.
 The AR(1) structure provided the best fit
according to BIC, and by using it we were able to obtain a
100\% classification rate using LOOCV on the dataset.
\cite{doi:10.1080/10618600.2018.1476249} report  a 90\% correct
classification rate using LOOCV on this dataset.

  \hypertarget{conclusion}{%
\section{Conclusions}\label{conclusion}}
We have provided an ECME method for fitting the parameters of the
MxV$t$ distribution that can be used on three-way data sets such as
multivariate repeated measures, image or spatial data, and have 
demonstrated the method on simulation datasets and on classification and
discrimination in four
real-world applications where the new method using the 
MXV$t$-distribution outperforms that using the MxVN.
The  ECME algorithm and the discriminant analysis are implemented  in  the \texttt{R} 
package \href{https://cran.r-project.org/package=MixMatrix}{\texttt{MixMatrix}}. 
The package also includes functions for sampling from and computing the density
of the MxVN and MxV$t$ distributions and includes the
datasets used in this paper.

Our model can be extended beyond supervised learning to mixture
 model-based clustering and can be made to accommodate
more specialized covariance structures such as those described in
\cite{fraley2002model} and \cite{teigen}. It may also be readily
extended to cases with incomplete records. 
Determining the existence, convergence and uniqueness properties would also be
desirable. For instance, we know how many observations are required
to have unique ML estimates of the parameters in the MxVN distribution with
unconstrained mean and  covariance matrices but such results may be
useful to develop for the MxV$t$ or the constrained MxVN. 
Nevertheless, the EM algorithm is guaranteed to converge to a local
stationary point, provided it is initialized where  the
log-likelihood function is finite~\cite{wu1983}. Finally, another
area that could benefit from further development is 
the extension of {\tt MatrixLDA} to include the MxV$t$ distribution,
where we believe our development in this paper will be helpful.
\section*{Funding Information}
      This research was supported in part by the
      National Institute of Justice (NIJ) under Grants
      No. 2015-DN-BX-K056 and 2018-R2-CX-0034. The research of the
      second author was also supported in part by the National
      Institute of Biomedical Imaging and Bioengineering (NIBIB) 
of the National Institutes of Health (NIH) under Grant R21EB016212,
and the United States Department of Agriculture (USDA) National
Institute of Food and Agriculture (NIFA) Hatch project IOW03617.
The content of this paper is however solely the responsibility of the
authors and does not represent the official views of the NIJ, the
NIBIB, the NIH, the NIFA or the USDA.


\ifCLASSOPTIONcaptionsoff
  \newpage
\fi

\bibliographystyle{IEEEtran}

\bibliography{matrixpaper}
\newpage
\renewcommand\thefigure{S-\arabic{figure}}\setcounter{figure}{0}
\renewcommand\thetable{S-\arabic{table}}
\renewcommand\thesection{S-\arabic{section}}
\renewcommand\thesubsection{S-\arabic{section}.\arabic{subsection}}
\renewcommand\theequation{S-\arabic{equation}}

\section*{Supplementary Materials}
\setlength{\tabcolsep}{1pt}

\section{The EM algorithm for parameter estimation in the MxV$t$ distribution}
\label{em}
As mentioned in the paper, the MxV$t$ distribution does not
have closed-form ML estimators so we develop an EM algorithm by
augmenting the data, in similar spirit as done for the vector-multivariate
$t$-distribution~\cite{lange1989robust}, and then present an ECME (Expectation/Conditional Maximization Either) algorithm~\cite{liurubin1994}
to improve the speed of convergence of the EM algorithm. Let $\bX_1,\bX_2,\ldots,\bX_n$ be
independent realizations from the $t_{p,q}(\nu,\bfM,\bSigma,\bOmega)$
density. Then each $\bX_i$ can be augmented with latent Wishart-distributed weight
matrices $\bS_i$ as follows:
\begin{equation}
  \begin{split}
  \bX_i | \mathbf{M}, \bSigma, \bOmega, \nu, \bS_i & \sim   {\mathcal N}_{p,q}(\mathbf{M}, \bS_i^{-1}, \bOmega) \\
  \bS_i | \mathbf{M}, \bOmega, \bSigma, \nu & \sim  \mathcal{W}_p(\nu + p -1, \bSigma^{-1}),  \quad \mathrm{for } \,\, i = 1, 2, \ldots, n .
\end{split}
  \label{eqn:sdefinition}
\end{equation}
To show the benefits of using the latent $\bS_i$s, we first derive ML
estimators with the complete data and then use that to derive an EM 
algorithm using only the observed data. We then modify the EM
algorithm to its more efficient ECME derivative. 

\hypertarget{ml-estimation-of-parameters-with-observed-x-and-s.}{%
\subsection{ML Estimation of parameters with complete data}}\label{ml-estimation-of-parameters-with-observed-x-and-s.}

Suppose that we have $(\bX_i,\bS_i), i=1,2,\ldots,n$ where each
$\bS_i\sim {\mathcal W}_p(\nu + p -1, \bSigma^{-1})$ and  $\bX_i\mid
\bS_i\sim {\mathcal N}_{p,q}(\bfM,\bS^{-1}, \bOmega)$ for each
$i=1,2,\ldots,n$. Then the complete log-likelihood function $\ell_c$ of the parameters
$(\mathbf{M}, \bOmega)$ given the data $(\bX_i,\bS_i), 
i=1,2,\ldots,n$ can written as a sum of (conditional) MxVN
log-likelihood functions $\ell_N$ and a sum of Wishart log-likelihood
functions $\ell_W$:

\begin{align*}\ell_c(\mathbf{M},\bSigma,\bOmega,\nu;\bX,\bS) &= \ell_N(\mathbf{M},\bS^{-1} , \bOmega ; \bX\mid \bS) + \ell_W(\nu,\bSigma ; \bS)\end{align*}
From the definitions of the MxVN and Wishart distributions, we have,
after ignoring  additive constants, 
\begin{equation*}
\begin{split}
  \ell_N (\mathbf{M},\bS^{-1} , \bOmega ; \bX\mid \bS)  
  &= -\frac{np}{2} \log | \bOmega| + \frac{q}{2}\sum_{i = 1}^n \log|\bS_i| \\
  & \quad - \frac{1}{2}\mathrm{tr}\left[ \sum_{i = 1}^n{\bS_i \bX_i
      \bOmega^{-1}\bX_i^T} + (\sum_{i = 1}^n{\bS_i})\mathbf{M}
    \bOmega^{-1}\mathbf{M}^T - 2 (\sum_{i =
      1}^n{\bS_i\bX_i})\bOmega^{-1}\mathbf{M}^T  \right] \\
\end{split}
\end{equation*}
and 
\begin{equation*}
\begin{split}
\ell_W(\nu,\bSigma ; \bS) &=  (\nu-2)/2 \sum_{i = 1}^n\log |\bS_i| - \sum_{i = 1}^n \mathrm{tr}(\bSigma \bS_i )/2 - n\nu p / 2 \log 2 \\
                   & \quad +
                     n (\nu + p - 1)/2 \log|\bSigma | - n\log\Gamma
                     _{p}( (\nu + p -1)/2).
                     \end{split}
                   \end{equation*}
To simplify computation of the ML estimators and their notation, we define the
following complete data sufficient statistics for the parameters: 
\begin{align*} \bS_{SX} =\sum_{i = 1}^n \bS_i \bX_i; \quad \bS_S =  \sum_{i = 1}^n \bS_i; \quad \bS_{XSX} =  \sum_{i = 1}^n  \bX_i^T \bS_i \bX_i \quad \bS_{|S|} = \sum_{i = 1}^n \log |\bS_i|. \end{align*}

Taking derivatives of log-likelihoods yields the  ML estimates:
\begin{align*}\widehat{\mathbf{M}} &= \left(\sum_{i = 1}^n\bS_i\right)^{-1}\sum_{i=1}^n\bS_i\bX_i  = \bS_{S}^{-1} \bS_{SX}, \\
\widehat{\bOmega} &= \frac{1}{np} \sum_{i = 1}^n (\bX_i - \widehat{\mathbf{M}})^T \bS_i (\bX_i - \widehat{\mathbf{M}}) = \frac{1}{np} \left(\bS_{XSX} - \bS_{SX}^T \bS_{S}^{-1}\bS_{SX}  \right),\\
\widehat{\bSigma}^{-1} &= \frac{1}{n(\nu+p-1)}\sum_{i = 1}^n \bS_i = \frac{\bS_S}{n(\nu+p-1)}.
\end{align*}

The ML estimate of $\nu$ can be obtained by finding the root of the equation:
\[ n \psi_p ((\nu+p-1)/2) - (\bS_{|S|} - np\log2 + n \log |\bSigma|) = 0\]
with $\psi_p(\cdot)$ the $p$-variate digamma function, defined as
$\psi_p(x) = d \log\Gamma_p(x)/dx$. The ML estimate of $\nu$ may be obtained
numerically by a one-dimensional search algorithm. We now use the
development in this section in our EM algorithm for a sample from the
MxV$t$ distribution. 

\subsection{Estimating parameters from a MxV$t$ sample}\label{mle-of-parameters-m-omega-sigma-and-nu-using-ecme}

\subsubsection{The EM algorithm}
 \label{s-em}

Let $\bX_i ,i=1,2,\ldots,n$ be independent identically distributed 
realizations from $t_{p,q}(\nu,\bfM,\bSigma,\bOmega)$. As in the main
article, we write $\bTheta\equiv\{\nu,\bfM,\bSigma,\bOmega\}$.
From the development in the introduction of this section, for each
$i=1,2,\ldots,n$, let $\bS_i$ be (unobserved) random matrices as per 
Equation~\eqref{eqn:sdefinition} and Property~\ref{prop2}. Then the expected complete log-likelihood
function is
\begin{equation}
  \begin{split}
Q(\Theta; \Theta^{(t)}) &= -\frac{np}{2} \log | \bOmega|  - \frac{n\nu p\log 2 }{  2} - n\log\Gamma _{p}\left(\frac{\nu + p -1}{2}\right)+ n \frac{\nu + p - 1}{2} \log|\bSigma |  \\
  & \quad +\E_{\Theta^{(t)}}\left\{\left[ - \frac{1}{2}\mathrm{tr}\left( \sum_{i = 1}^n{\bS_i \bX_i \bOmega^{-1}\bX_i^T} + \sum_{i = 1}^n{\bS_i}\mathbf{M} \bOmega^{-1}\mathbf{M}^T - 2 \sum_{i = 1}^n{\bS_i\bX_i}\bOmega^{-1}\mathbf{M}^T  \right)\right.\right. \\
  &\qquad\qquad \left.\left.+  \frac{\nu-2}2 \sum_{i = 1}^n\log |\bS_i| -
\frac12\sum_{i = 1}^n \mathrm{tr}(\bSigma \bS_i )   +
\frac{q}{2}\sum_{i = 1}^n \log|\bS_i| \bigg| \bX_1,\bX_2,\ldots,\bX_n \right]\right\}.
  \end{split}
  \label{qfn}
\end{equation}
\paragraph{E-step}\label{e-step}
Using Property~\ref{prop2} and properties of the Wishart distribution, the 
expectation step (E-step) updates 
at the 
current value  $\bTheta^{(t)}$ of $\bTheta$ are,  by taking the expected values
of the $\bS_i$ given the current value of  $\bTheta^{(t)}$:
\begin{equation*}
  \begin{split}
\bS_{i}^{(t+1)} \doteq\E_{\Theta^{(t)}}(\bS_i | \bX_i) &= (\nu^{(t)}+p+q-1)[(\bX_i-\mathbf{M}^{(t)})\bOmega^{(t)^{-1 }}(\bX_i-\mathbf{M}^{(t)})^T + \bSigma^{(t)}]^{-1}, \\
\E_{\Theta^{(t)}}\left(\log| \bS_i| \big| \bX_i \right)   &=\psi_{p}\left({\frac {\nu^{(t)} + p + q -1}{2}}\right) + p \log 2 + \log \left|\frac{\bS_{i}^{(t+1)}}{\nu^{(t)} + p + q -1}\right|,
\end{split}
\end{equation*}
with $\psi_p(\cdot)$ as the $p$-variate digamma function.
Note that the updates for $\bS_i^{(t+1)}$ exist by construction
if the $\bSigma$ and $\bOmega$ are positive definite. We define
and store the expected sufficient statistics to reduce computational calculations
and for notational convenience:
\begin{equation*}
  \begin{split}
    \bS_{S}^{(t+1)} &\doteq \sum_{i=1}^n \bS_{i}^{(t+1)},\\
    \bS_{SX}^{(t+1)} & \doteq  \sum_{i=1}^n \E_{\Theta^{(t)}} (\bS_i\bX_i|\bX_i)
=\sum_{i=1}^n \bS_{i}^{(t+1)}\bX_i, \\
\bS_{XSX}^{(t+1)} &\doteq \sum_{i=1}^n\E_{\Theta^{(t)}}
(\bX_i^T\bS_i\bX_i|\bX_i) = \sum_{i=1}^n
\bX_i^T\bS_{i}^{(t+1)}\bX_i,\\
\bS_{|S|}^{(t+1)} & \doteq \E_{\Theta^{(t)}}  \left[\sum_{i = 1}^n \log| \bS_i|  \bigg| \bX_i \right], 
\end{split}
\end{equation*}
with the last expression needed only when we are also estimating $\nu$. 
In that case, these statistics can be expressed with $(\nu^{(t)} + p +
q -1)$ 
factored out, and for convenience may be computed and stored that way
when $\nu$  needs to be estimated. These quantities can be computed
in $\bOO(npq^2) + \bOO(np^2q) + \bOO(np^3)$ flops.

\paragraph{Maximization step}\label{maximization-step}
Based on the updated weight matrices $\bS_i^{(t+1)}$ and statistics
based on $\Theta^{(t)}$ and $\bX$, we get the updates:
\begin{align*}
  \widehat{\mathbf{M}} &= \left(\sum_{i = 1}^n\bS_i^{(t+1)}\right)^{-1}\sum_{i=1}^n\bS_i\bX_i  = {\bS_{S}^{(t+1)}}^{-1} \bS_{SX}^{(t+1)} ,\\
  \widehat{\bOmega} &= \frac{1}{np} \sum_{i = 1}^n (\bX_i - {\mathbf{M}}^{(t)})^T \bS_i^{(t+1)} (\bX_i - {\mathbf{M}}^{(t)}) \\
                     &=\frac{1}{np} \left(\bS_{XSX}^{(t+1)} - {\bS_{SX}^{(t+1)}}^T {\bS_{S}^{(t+1)}}^{-1}{\bS_{SX}^{(t+1)}}  \right), \\
 \widehat{\bSigma}^{-1} &= \frac{1}{n(\nu^{(t)}+p-1)}\sum_{i = 1}^n \bS_i^{(t+1)} = \frac{\bS_S^{(t+1)}}{n(\nu^{(t)}+p-1)}.
\end{align*}
This can be computed in $\bOO(p^2q) + \bOO(pq^2)$ flops, which is negligible compared to the E-step computations.
Again, treating the set of $\bS_i^{(t+1)}$ as observed, the MLE of
$\nu$ can be obtained:
\[ n \frac{d}{d\nu}\log \Gamma_p ((\nu+p-1)/2) - \frac{1}{2} (\bS_{|S|} - np\log2 + n \log |\widehat{\bSigma}|) = 0.\]
Defining $\kappa^{(t)} = \nu^{(t)} +p + q -1$ for compactness: \begin{align}
0 &= n \psi_p((\nu+p-1)/2) -  \left(n\psi_{p}\left({\frac {\kappa^{(t})}{2}}\right)  +\sum_{i=1}^n \log \left|\frac{\bS_{i}^{(t+1)}}{\kappa^{(t)}}\right| - n \log \left|\frac{\bS_S^{(t+1)}}{n(\nu^{(t)}+p-1)}\right|\right)  \nonumber \\
 &=  \psi_p((\nu+p-1)/2) - \left(\psi_{p}\left({\frac {\kappa^{(t)}}{2}}\right)  +\frac{1}{n}\sum_{i=1}^n \log \left|\bZ_{i}^{(t+1)}\right| + p \log \frac{n(\nu^{(t)} + p -1)}{\kappa^{(t)}} - \log \left|\bZ_S^{(t+1)}\right|\right) 
\label{eqn:ecm}\end{align} where $\bZ_{*}$ is the appropriate $\bS_{*}$ statistic with
$(\nu^{(t)} + p + q -1)$ factored out and $\psi_p$ is the
$p$-dimensional digamma function. This can be solved using
a 1-dimensional search.

Since each $\bS_i$ is positive definite by construction if the
previous $\bSigma^{(t)}$ and $\bOmega^{(t)}$ were positive definite,
the updates $\widehat{\bSigma}$ and $\widehat{\mathbf{M}}$ exist. The conditions for
the positive definiteness of the update $\widehat{\bOmega}$
are less clear: it is the sum of matrices only guaranteed to be
positive semi-definite and we do not have a proof of the necessary or sufficient
sample size to guarantee the update is positive definite (a.s.) as required
for the method. A solution for $\hat{\nu}$ is guaranteed to exist as long
as $\widehat{\bSigma}$ and $\widehat{\bOmega}$ exist and are positive
definite.

\paragraph{ML Estimation with the Expectation/Conditional Maximization Either (ECME) algorithm}

First we note that, if $\nu$ is known, there is no need to partition
the M-step into multiple constrained maximization steps. If $\nu$ is
required to be estimated, there is no difference
between a standard EM and a standard ECM (Expectation/Conditional Maximization) algorithm
in this setting, since, as in the case of the multivariate $t$ distribution,
the complete data likelihood function factorizes into  $\Theta_1 = (\mathbf{M}, \bSigma, \bOmega)$
and  $\Theta_2 = (\nu)$. However, by partitioning it in this way, it is
possible, similarly to the case of the multivariate $t$, to find a more
efficient method of maximization. This is desirable because the M-step for
$\nu$ can be slow.  Here we present an ECME (Expectation/Conditional Maximization Either)
algorithm that first maximizes the expected log-likelihood for
$(\mathbf{M}, \bSigma, \bOmega)$ and then maximizes the actual log-likelihood
over $\nu$ given the current values $(\mathbf{M}, \bSigma, \bOmega)$, similar to \cite{liurubin1994}.

Given  $\Theta_1 = (\mathbf{M}, \bSigma, \bOmega)$, we can maximize for $\nu$
in Equation~\eqref{mleqn}, yielding the  set of equations provided in \eqref{eqn:ecme}
\begin{align*}
 0 &= n \psi_p((\nu+p-1)/2) -  \left\{n\psi_{p}\left({\frac {\kappa}{2}}\right)  +\sum_{i=1}^n \log \left|\frac{\bS_{i}^{(t+1)}}{\kappa}\right| - n \log \left|\frac{\bS_S^{(t+1)}}{n(\nu+p-1)}\right|\right\} \nonumber \\
 &=  \psi_p((\nu+p-1)/2) - \left\{\psi_{p}\left({\frac {\kappa}{2}}\right)  +\frac{1}{n}\sum_{i=1}^n \log \left|\bZ_{i}^{(t+1)}\right| + p \log \frac{n(\nu + p -1)}{\kappa} - \log \left|\bZ_S^{(t+1)}\right|\right\}.
\label{eqn:ecme}\end{align*}
The difference is that the solution for $\nu^{(t+1)}$ no longer depends on $\nu^{(t)}$,
Solving this equation is slightly more computationally complex  than solving
Equation~\eqref{eqn:ecm} ($\nu$ appears four times in the equation to be solved
rather than once) but this converges in fewer total iterations.
The ML estimating equation can be solved by a one-dimensional
search, providing a ECME algorithm with the steps (as also provided in
the main article):
\begin{enumerate}
\def\labelenumi{\arabic{enumi}.}
\item
  {\bf E-step:} Update $\bS_i$ weights and statistics based on
  $\Theta^{(t)}$ and $\bX$.
\item
  {\bf CME-step:} Update
  $\Theta_1^{(t+1)} = (\mathbf{M}^{(t+1)}, \bSigma^{(t+1)}, \bOmega^{(t+1)})$.
\item
  {\bf CME-step:} Update $\Theta_2^{(t+1)} = \nu^{(t+1)}$ using the observed
  log-likelihood given the current values $(\mathbf{M}^{(t+1)},\bSigma^{(t+1)},\bOmega^{(t+1)})$ by solving Equation~\eqref{eqn:ecme}.
\end{enumerate}
Repeat these steps until convergence. Each iteration of this algorithm takes
$\bOO(npq^2) + \bOO(np^2q) + \bOO(np^3)$ flops plus the number of
iterations required by the second CME step.

\subsubsection{Fitting with restrictions on the
parameters}\label{fitting-with-restrictions-on-the-parameters}

In some settings, restrictions on the parametrization of the center or
scatter matrices are appropriate. In this section, we derive
solutions in the cases of center matrices that are constant across rows,
columns, or the entire matrix. In \cite{ROY2005462} some results
for restrictions on covariance matrices were derived and in this paper
AR(1) covariance structures and compound symmetry (CS) variance
structures were used; however, they were fit numerically as closed forms
for the derivatives and determinants exist. Let $\mathbf{1}_{p,q}$
denote a $(p \times q)$ matrix consisting only of $1$s. Then it can
be shown that these are the appropriate M-step estimates for certain
mean matrix constraints: \begin{align*}
\mathbf{M} &= \mathbf{1}_{p,q} \mu: & \widehat{\mathbf{M}} &= \mathrm{tr}(\bS_{SX} \widehat{\bOmega}^{-1}\mathbf{1}_{q,p})/\mathrm{tr}(\bS_S \mathbf{1}_{p,q}  \widehat{\bOmega}^{-1}\mathbf{1}_{q,p})  \mathbf{1}_{p,q}\\
\mathbf{M} &= \mathbf{1}_{p,1} \mathbf{\mu}_{1,q}: & \widehat{\mathbf{M}} &= \mathbf{1}_{p,p} \bS_{SX} / (\mathbf{1}_{1,p}\bS_S\mathbf{1}_{p,1})\\
\mathbf{M} &=  \mathbf{\mu}_{p,1} \mathbf{1}_{1,q}: & \widehat{\mathbf{M}} &= \bS_S^{-1} \bS_{SX} \widehat{\bOmega}^{-1} \mathbf{1}_{q,q} / (\mathbf{1}_{1,q}\widehat{\bOmega}^{-1}\mathbf{1}_{q,1})
\end{align*}
which can be used to simplify the ECME algorithms further.

\subsection{Performance Evaluations}\label{performance-evaluations}
\subsubsection{Simulation Study}\label{supp:simulation-study}
\begin{figure}
\centering
    \mbox{
      \subfloat[]{\includegraphics[width=.5\textwidth]{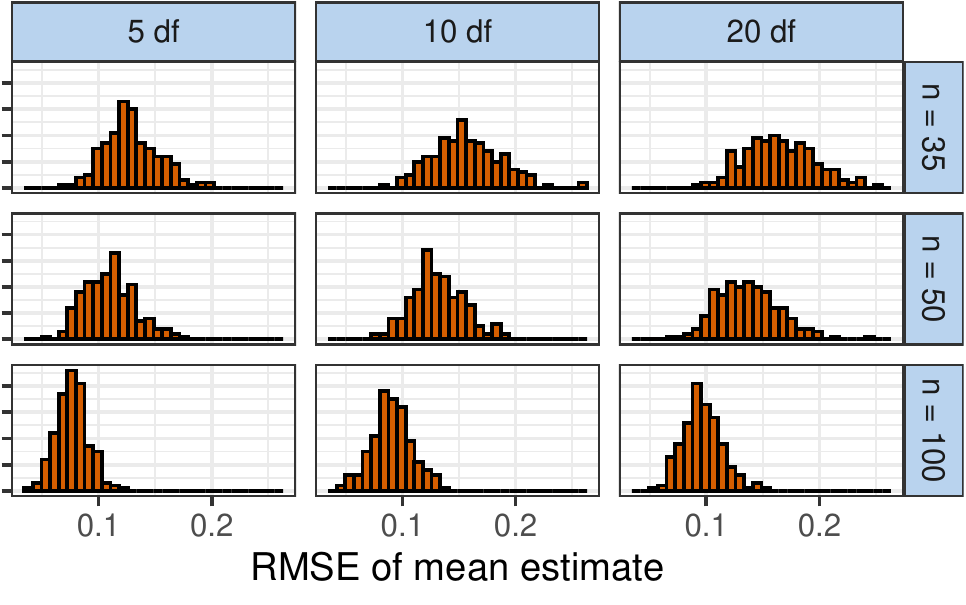}}
  \subfloat[]{{\includegraphics[width=.5\textwidth]{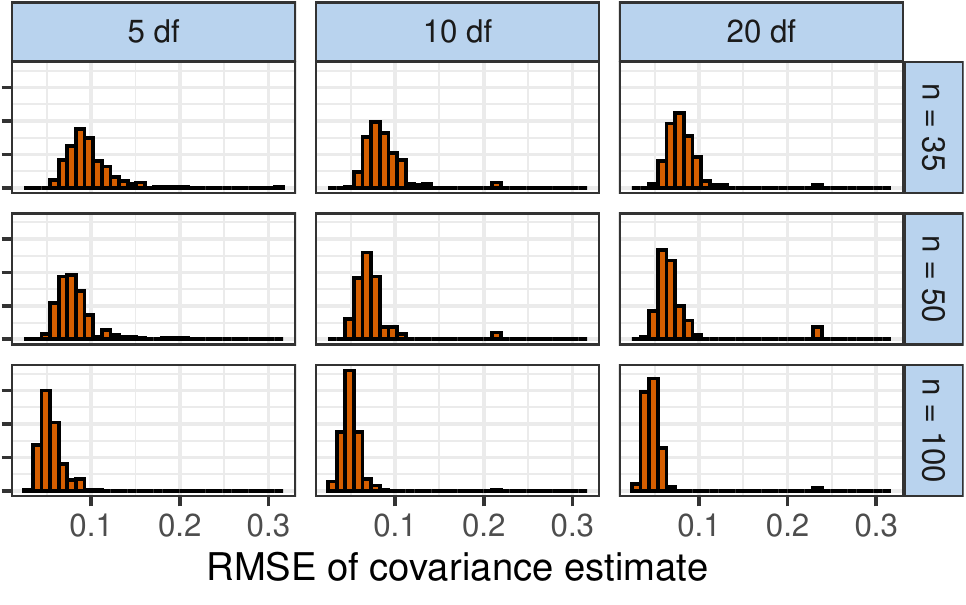}}}}
\caption{(a) RMSE for the mean estimates and (b) RMSE for the covariance estimates
  for datasets of size $n=35, 50, 100$ with true $\nu=5, 10, 20$.\vspace{0.1in} }
\label{fig:rmse}
\vspace{-.2in}
\end{figure}
In the main paper, results pertaining to the recovery of the $\nu$ parameter
were reported for a simulation study where 200 datasets were produced for
$\nu = 5, 10, 20$ and $n = 35, 50, 100$ with a 0 mean matrix and
identity scatter matrices. Here we report also the
results for the recovery of the mean and covariance parameters.
For $\bX \sim t(\nu, \bM, \bSigma, \bOmega)$, we have the result that
$\mathrm{cov}(\mathrm{vec}(\bX)) = \bSigma \otimes \bOmega / (\nu - 2)$.
To compare all nine sets of simulations on the same scale, we correct each by the appropriate
scaling factor such that each has an identity covariance matrix and then
report the root mean square difference between the actual and fitted
$\hat{\bM}$ and $\hat{\bSigma} \otimes \hat{\bOmega}$ in Figure~\ref{fig:rmse}.
The figures indicate performance improves as the sample size increases and indicates
good recovery of the parameters in every case. 

\begin{wrapfigure}{r}{.75\textwidth}
  \centering
  \vspace{-.1in}
\includegraphics[width = .75\textwidth]{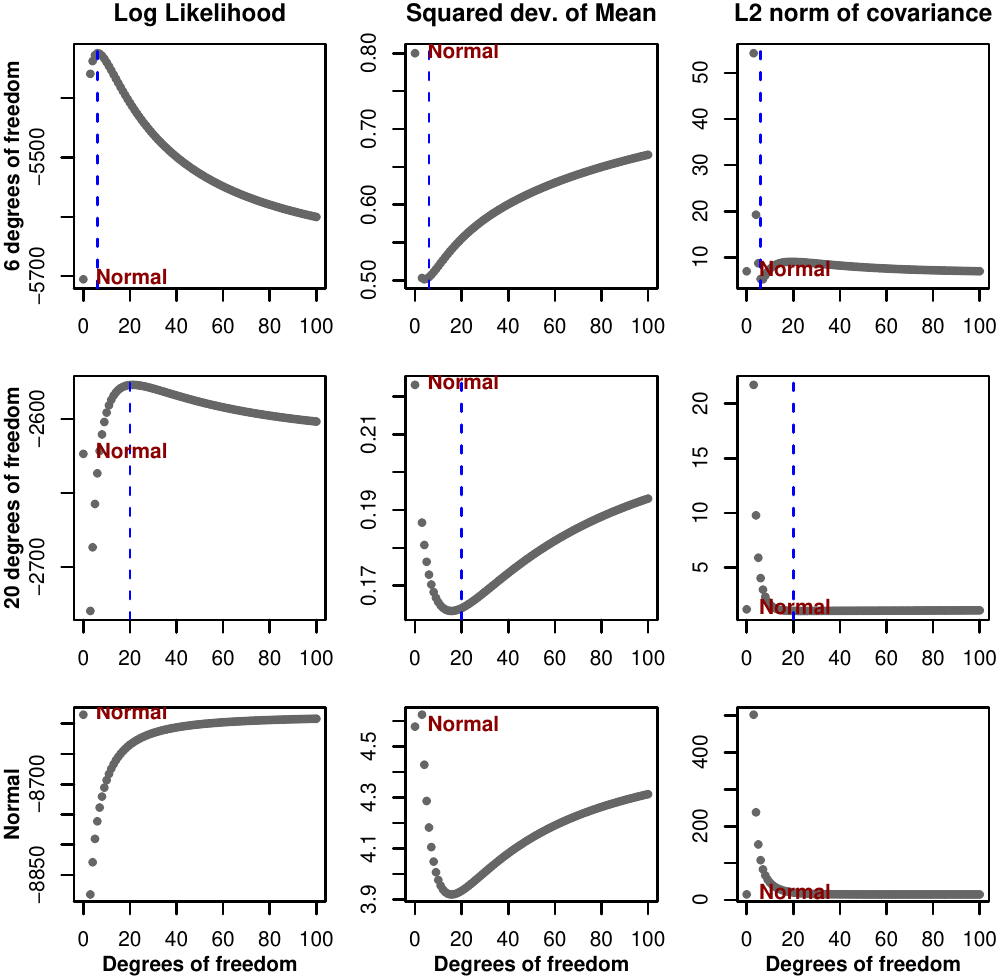}
\vspace{-0.2in}
 \caption{The top rows contains results for a true $\nu = 6$ and $20$ (the blue line)
  while the bottom row contains the results for a true matrix normal distribution.
    \vspace{0.2in}
  }\label{fig:suppstudy}
\vspace{-.4in}
\end{wrapfigure}
We provide a second simulation study to address concerns about model misspecification,
namely, what happens when a matrix $t$ distribution model is treated as a matrix
normal or vice versa. Three datasets of size 100 with mean matrix 0 and parameters
$\bSigma$ a $5 \times 5$ AR(1) matrix with $\rho = 0.7$ and $\bOmega$ a draw from a
standard Wishart distribution with $\nu = 10$ and dimension 8, with one dataset from a
MxV$t$ distribution with 6 degrees of freedom, one with 20 degrees of freedom,
and one from a MxVN distribution.

In Figure~\ref{fig:suppstudy}, we plot the log-likelihood, squared deviation from the mean,
and the $L^2$ distance between the true and estimated covariance matrix. The
top two rows indicate the results for the MxV$t$ with $\nu = 6$ and $20$ and the bottom indicates the
results for the MxVN, fitted to a MxVN and to MxV$t$ models with $\nu = 3, 4, \ldots, 100$.
On the MxV$t$ with $\nu = 6$ and $20$,
the MxVN performed poorly compared to the MxV$t$ with $\nu$ near the true parameter values.
On the MxVN, the MxV$t$ performed poorly. 

For all of the datasets, the MxVN has slightly worse recovery of the mean
matrix than the MxV$t$ distributions while the MxVN had estimates of the
covariance matrix that were comparable to the best MxV$t$ estimates. The $L^2$
norm of the covariance matrix was not accurate for low values of $\nu$.

The behavior here is suggestive of what occurs in the results when the
method fails to converge. Simulations that fail to converge slowly
increase likelihood as
$\nu$ increases until either the maximum number of iterations or the upper
bound of $\nu$ is reached. This scenario occurs more frequently when
simulating from distributions with large $\nu$ and small sample sizes
or simulating from MxVN distributions with modest sample sizes (for
larger sample sizes, even an MxVN will usually converge to some
distribution with large $\nu$). As Figure~\ref{fig:suppstudy} indicates, the
likelihood surface is very flat for a true MxVN across values of $\nu$.
With a small sample size and $\nu$ not small, this may occur there was well.

\subsubsection{Matching Fractured Surfaces}\label{supp-matching-of-fractured-surfaces}

\begin{wrapfigure}{r}{.7\textwidth}
  \centering
\includegraphics[width = .7\textwidth]{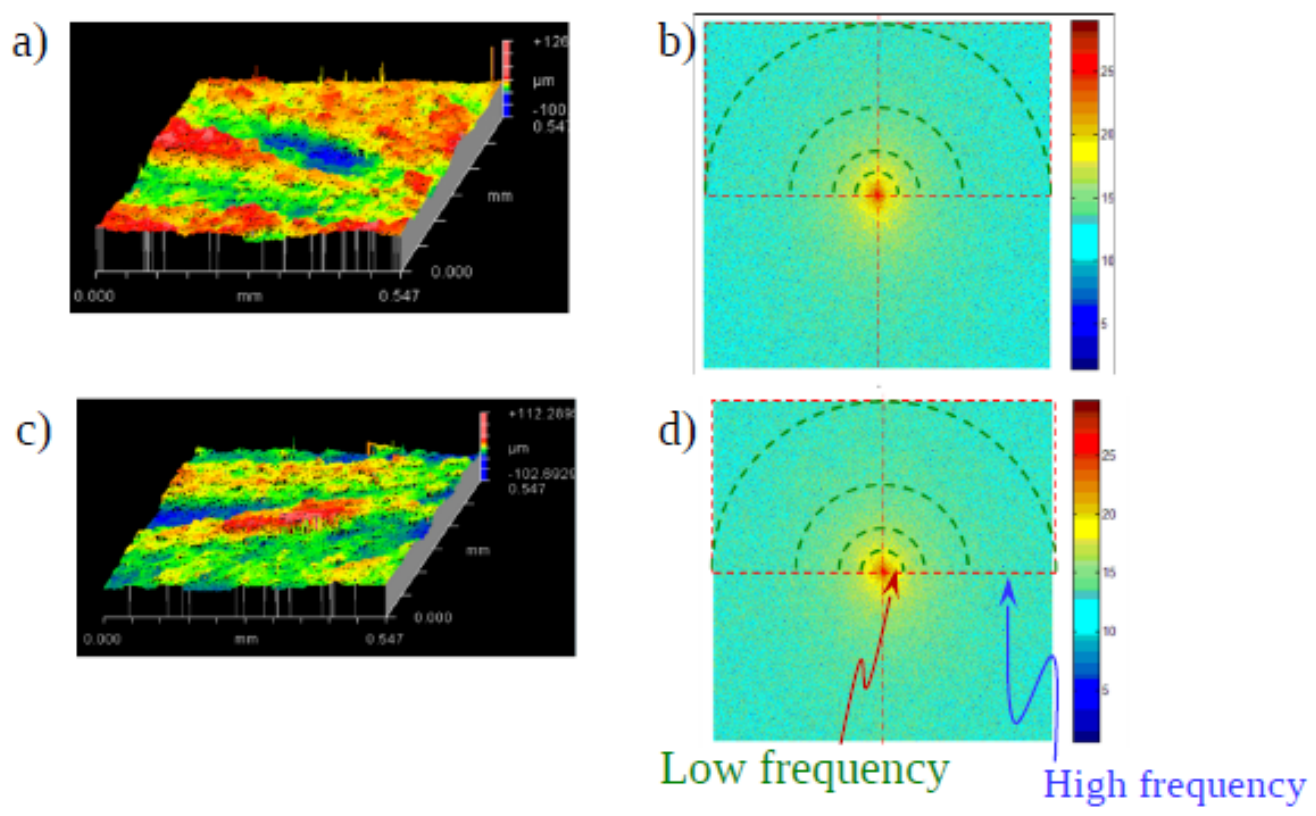}
\singlespace \caption{ Surface height 3D topographic maps for
  tip and base pair (a,c) and their corresponding 2D spectral
  analysis (b,d).
    \vspace{0.1in}
  }\label{fig:knifeimg}
\end{wrapfigure}
The knife surfaces were scanned using a standard non-contact
3D optical interferometer in corresponding regions, then the 2D
Fourier frequencies were computed and compared. In Figure~\ref{fig:knifeimg},
we illustrate one pair of corresponding images (out of 9) from one of the
knife base-tip pairs (out of 38). On the left are a visualization of the output
of the 3D optical interferometer for the two surfaces. Note that the
images are presented as-is - they should fit together when one is
flipped over. The blue depressed region on the top corresponds to the red
elevated region on the bottom. On the right is a visualization of the 2D
Fourier transform with the frequency ranges used for comparison
highlighted - the two bands between the ``low frequency'' and ``high
frequency'' region. 

\end{document}